\documentstyle{article}
\begin{document}
  \title{$S_4\times Z_2$ Flavor Symmetry in  Supersymmetric Extra U(1) Model}
  \author{Y. Daikoku\footnote{E-mail: yasu\_daikoku@yahoo.co.jp} \quad and \quad H. Okada\footnote{E-mail: HOkada@bue.edu.eg} \\
  {\em Institute for Theoretical Physics, Kanazawa University, Kanazawa 920-1192, Japan.}$^*$ \\
  {\em Centre for Theoretical Physics, The British University in Egypt.}$^\dagger$}
  \maketitle
\begin{abstract}
 We propose a $E_6$ inspired supersymmetric model with a non-Abelian discrete flavor symmetry ($S_4$ group); 
that is, $SU(3)_c\times SU(2)_W\times U(1)_Y \times U(1)_X\times S_4\times Z_2$.
 In our scenario, the additional abelian gauge symmetry; $U(1)_X$, not only solves the $\mu$-problem in the minimal Supersymmetric Standard Model(MSSM), but also requires new exotic fields which play an important role in solving flavor puzzles.
 If our exotic quarks can be embedded into a $S_4$ triplet, which corresponds to the number of the generation, one finds that dangerous proton decay can be well-suppressed.
Hence, it might be expected that the generation structure for lepton and quark in the SM(Standard Model) can be understood as a new system in order to 
stabilize the proton in a supersymemtric standard model (SUSY).

 Moreover, due to the nature of the discrete non-Abelian symmetry itself,  Yukawa coupling constants of our model are drastically reduced. In our paper, 
we show two predictive examples of the models for quark sector and lepton sector, respectively.

\end{abstract}

\newpage

\def\vol#1#2#3{{\bf {#1}} ({#2}) {#3}}
\def\NP{Nucl.~Phys. }
\def\PL{Phys.~Lett. }
\def\PR{Phys.~Rev. }
\def\PRP{Phys.~Rep. }
\def\PRL{Phys.~Rev.~Lett. }
\def\PTP{Prog.~Theor.~Phys. }
\def\MPL{Mod.~Phys.~Lett. }
\def\IJMP{Int.~J.~Mod.~Phys. }
\def\JETP{Sov.~Phys.~JETP }
\def\JP{J.~Phys. }
\def\ZP{Z.~Phys. }
\def\EPJ{Eur.~Phys.~J.}
\def\th#1{hep-th/{#1}}
\def\ph#1{hep-ph/{#1}}
\def\hex#1{hep-ex/{#1}}
\def\PAN{Phys.~Atomic.~Nucl.}
\def\AST{Astrophys.~J.}

\def\equ#1{\begin{equation}
#1
\end{equation}
}
\def\nn{\nonumber}
\def\no{\nonumber}
\def\rvec#1{\overrightarrow{#1}}
\def\lvec#1{\overleftarrow{#1}}

\def\2tvec#1#2{
\left(
\begin{array}{c}
#1  \\
#2  \\   
\end{array}
\right)}

\def\mat2#1#2#3#4{
\left(
\begin{array}{cc}
#1 & #2 \\
#3 & #4 \\
\end{array}
\right)
}

\def\Mat3#1#2#3#4#5#6#7#8#9{
\left(
\begin{array}{ccc}
#1 & #2 & #3 \\
#4 & #5 & #6 \\
#7 & #8 & #9 \\
\end{array}
\right)
}

\def\3tvec#1#2#3{
\left(
\begin{array}{c}
#1  \\
#2  \\   
#3  \\
\end{array}
\right)}

\def\4tvec#1#2#3#4{
\left(
\begin{array}{c}
#1  \\
#2  \\   
#3  \\
#4  \\
\end{array}
\right)}

\def\L{\left}
\def\R{\right}

\def\pl{\partial}

\def\lra{\leftrightarrow}

\def\hbar{\hspace{1mm}\bar{}\hspace{-1mm}h}

\def\eqn#1{
\begin{eqnarray}
#1
\end{eqnarray}
}

\def\eqno#1{
\begin{eqnarray*}
#1
\end{eqnarray*}
}

\def\etensor{\epsilon_{\mu\nu\rho\sigma}}

\def\class#1#2#3#4{
\left\{
\begin{array}{ll}
\displaystyle{#1} &
  \qquad #2 \\[0.2cm]
\displaystyle{#3} &
  \qquad #4
\end{array}
\right.
}

\def\classt#1#2#3#4#5#6{
\left\{
\begin{array}{ll}
\displaystyle{#1} &
  \qquad #2 \\[0.2cm]
\displaystyle{#3} &
  \qquad #4 \\[0.2cm]
\displaystyle{#5} &
  \qquad #6
\end{array}
\right.
}

\def\therefore{
\raisebox{.2ex}{.}
\raisebox{1.2ex}{.}
\raisebox{.2ex}{.}
}

\def\g{{\rm g}}

\def\dbar#1{\overline{\overline{#1}}}

\def\dsl#1{#1\hspace{-2.1mm}\slash}

\section{Introduction}

 It is well-known that the standard model 
based on $G_{SM}=SU(3)_c\times SU(2)_W\times U(1)_Y$ gauge symmetry 
is a quite promising theory to describe interactions of the particles.

However,
there are unsolved or non-verifiable points enough, 
in particular, the followings are underlying to be clarified:
\begin{enumerate}
\item The electroweak symmetry breaking scale $M_W\sim 10^2$ ${\rm GeV}$ is unnaturally small 
in comparison with the fundamental energy scale such as Planck scale $M_P\sim 10^{18}$ {\rm GeV}.
\item  The number of Yukawa coupling constants is too many 
to give predictions of the quark and lepton mass matrices.
\item  There is no understanding about the meaning of generations.  
\end{enumerate}




It is believed that the first point 
is solved by introducing SUSY 
\cite{SUSY}, but there is still naturalness problem in the MSSM.
The superpotential of MSSM has $\mu$-term:
\eqn{
\mu H^U H^D.
}
The parameter $\mu$ has to be fine-tuned to $O(1\ {\rm TeV})$ in order to give appropriate electroweak breaking scale, 
 but it is unnatural.
This problem is elegantly solved by introducing an additional U(1) gauge symmetry.
This extra U(1) model is proposed in the context of superstring-inspired $E_6$ model \cite{extra-u1}.
In this model, the bare $\mu$-term is forbidden by the new $U(1)_X$ symmetry, but
the trilinear term including $G_{SM}$ singlet superfield $S$ is allowed:
\eqn{
\lambda SH^UH^D.
}
When this singlet field $S$ develops a vacuum expectation value (VEV),
the $U(1)_X$ gauge symmetry is spontaneously broken and an effective 
$\mu$-term; $\mu_{{\rm eff}}H^UH^D$, is generated from this term, where $\mu_{{\rm eff}}=\lambda \L<S\R>$ \cite{mu-problem}.

A promising solution for the second point 
is a flavor symmetry
\footnote{The $E_6$ inspired supersymmetric extension of SM with discrete flavor symmetry has been considered by authors \cite{f-extra-u1}.}.
In fact, the flavor symmetry strongly reduces the Yukawa coupling constants.
Here, we introduce a non-Abelian discrete flavor symmetry involved in triplet representations, 
expecting that the number of the generations for lepton and quark is {\it three}.  
The triplet representations are contained in
several non-abelian discrete symmetry groups \cite{review}, for examples,
$S_4$ \cite{s4}, $A_4$ \cite{a4}, $T'$ \cite{t-prime}, $\Delta(27)$ \cite{d27} and $\Delta(54)$ \cite{d54}.
In our work, we consider $S_4\times Z_2$.

A promising solution for the third point can be arose by 
the cooperation with the flavor symmetry and supersymmtery.
In the MSSM, the R-parity conserving operators such as $QQQL, E^cU^cU^cD^c$ induce the proton decay at unacceptable level. But, in the extra U(1) model,  these operators are forbidden by the additional gauge symmetry. 
However, since the extra U(1) model has additional exotic fields,
the Yukawa interactions for the exotic quarks and leptons and quarks reduce proton life time to unacceptable level, again.  
With the $S_4$ flavor symmetry, such a dangerous proton decay 
is sufficiently suppressed. 
Hence, it might be expected that the generation structure can be understood as a new system in order to 
{\em stabilize the proton}.  

The paper is organized as follows. 
In section 2, we explain the basic structure of $S_4$ flavor symmetric extra U(1) model.
We give the examples of the predictive model of lepton sector (model A) in section 3, 
and of quark sector (model B) in section 4.
In section 5, we discuss the flavon sector and its flavor violation.
Finally, we make a brief summary in section 6.
Experimental values of mixing matrices and masses of quarks and leptons are given in appendix,
which are used to test our models.

\section{The Extra U(1) Model with $S_4$ Flavor Symmetry}

\subsection{The Extra U(1) Model}

The basic structure of the extra U(1) model is given as follows. 
At high energy scale, the gauge symmetry of model has two extra U(1)s,
which consists maximal subgroup of $E_6$ as $G_2=G_{SM}\times U(1)_X\times U(1)_Z\subset E_6$.
MSSM superfields and additional superfields are embedded in three 27 multiplets of $E_6$ to cancel anomalies,
which is illustrated in Table 1.
The 27 multiplets are decomposed as ${\bf 27}\supset \L\{Q,U^c,E^c,D^c,L,N^c,H^D,g^c,H^U,\R.$ 
$\L.g,S\R\}$,
where $N^c$ are right-handed neutrinos (RHN), $g$ and $g^c$ are exotic quarks, and $S$ are  $G_{SM}$ singlets.
We introduce $G_{SM}\times U(1)_X$ singlets $\Phi$ and $\Phi^c$ to break $U(1)_Z$ 
which prevents the RHNs from having Majorana mass terms.
If the $G_{SM}\times U(1)_X$ singlets develop the intermediate scale VEVs along the D-flat direction
of $\L<\Phi\R>=\L<\Phi^c\R>$, then the $U(1)_Z$ is broken and the RHNs obtain the mass terms through
the trilinear terms $Y^M\Phi N^cN^c$ in the superpotential. After the symmetry is broken, as the R-parity symmetry
\eqn{
R=\exp\L[\frac{i\pi}{20}(3x-8y+15z)\R]
}
remains unbroken, $G_1=G_{SM}\times U(1)_X\times R$ survives at low energy.
This is the symmetry of the low energy extra U(1) model.

Within the renormalizable operators, the full $G_2$ symmetric superpotential is given as follows:
\eqn{
W_1&=&W_0+W_S+W_B, \\
W_0&=&Y^UH^UQU^c+Y^DH^DQD^c+Y^EH^DLE^c+Y^N H^ULN^c+Y^M\Phi N^cN^c, \\
W_S&=&kSgg^c+\lambda SH^UH^D, \\
W_B&=&\lambda_1 QQg+\lambda_2 g^cU^cD^c+\lambda_3 gE^cU^c+\lambda_4 g^cLQ+\lambda_5gD^cN^c.
}
For simplicity, we drop gauge and generation indices. 
Where $W_0$ is the same as the superpotential of the MSSM with the RHNs besides the absence of $\mu$-term, and
$W_S$ and $W_B$ are the new interactions.
In $W_S$, $kSgg^c$ drives the soft SUSY breaking
scalar squared mass of S to negative through the renormalization group equations and
then breaks $U(1)_X$ and generates mass terms of exotic quarks, and $\lambda SH^UH^D$ is source of the effective $\mu$-term.
Therefore, $W_0$ and $W_S$ are phenomenologically necessary.
In contrast, $W_B$ breaks baryon number and leads to very rapid proton decay, which are phenomenologically
unacceptable, so this must be forbidden.


\subsection{$S_4$ Flavor Symmetry}

We show how the $S_4$ flavor symmetry forbids the baryon number violating superpotential $W_B$.
Non-abelian group $S_4$ has two singlet representations ${\bf 1}$, ${\bf 1'}$, one doublet representation ${\bf 2}$ and two
triplet representations ${\bf 3}$, ${\bf 3'}$, where ${\bf 1}$ is the trivial representation.
As the generation number of quarks and leptons is three, 
at least one superfield of $\L\{Q,U^c,E^c,D^c,L,N^c,H^D,g^c,H^U,g,S\R\}$ must be assigned to triplet of $S_4$
in order to solve flavor puzzle. 
As we assume that full $E_6$ symmetry does not realize at Planck scale, there is no need to assign
all superfields to the same $S_4$ representations.
The multiplication rules of these representations are as follows:
\eqn{
\begin{tabular}{lcl}
${\bf 3}\times {\bf 3}={\bf 1}+{\bf 2}+{\bf 3}+{\bf 3'}$, & &
${\bf 3'}\times {\bf 3'}={\bf 1}+{\bf 2}+{\bf 3}+{\bf 3'}$, \\
${\bf 3}\times {\bf 3'}={\bf 1'}+{\bf 2}+{\bf 3}+{\bf 3'}$, & &
${\bf 2}\times {\bf 3}={\bf 3}+{\bf 3'}$, \\
${\bf 2}\times {\bf 3'}={\bf 3}+{\bf 3'}$, & &
${\bf 2}\times {\bf 2}={\bf 1}+{\bf 1'}+{\bf 2}$, \\
${\bf 1'}\times {\bf 3}={\bf 3'}$, & &
${\bf 1'}\times {\bf 3'}={\bf 3}$, \\
${\bf 1'}\times {\bf 2}={\bf 2}$, & &
${\bf 1'}\times {\bf 1'}={\bf 1}$.
\end{tabular}
}
With these rules, it is easily shown that
all the $S_4$ invariants consist of two or three non-trivial representations are given by
\eqn{
&&{\bf 1'}\cdot{\bf 1'},\quad {\bf 2}\cdot{\bf 2},\quad {\bf 3}\cdot{\bf 3},\quad {\bf 3'}\cdot{\bf 3'},\quad
{\bf 1'}\cdot{\bf 2}\cdot{\bf 2},\quad {\bf 1'}\cdot{\bf 3}\cdot{\bf 3'},\quad
{\bf 2}\cdot{\bf 2}\cdot{\bf 2}, \quad
{\bf 2}\cdot{\bf 3}\cdot{\bf 3}, \no \\
&&{\bf 2}\cdot{\bf 3}\cdot{\bf 3'},\quad
{\bf 2}\cdot{\bf 3'}\cdot{\bf 3'},\quad {\bf 3}\cdot{\bf 3}\cdot{\bf 3},\quad
{\bf 3}\cdot{\bf 3}\cdot{\bf 3'},\quad {\bf 3}\cdot{\bf 3'}\cdot{\bf 3'},\quad
{\bf 3'}\cdot{\bf 3'}\cdot{\bf 3'}.
}
From these, one can see that there is no invariant including only one triplet
\footnote{$T'$ does not have this property but $A_4$, $\Delta(27)$ and $\Delta(54)$ have.}.
Therefore, if $g$ and $g^c$ are assigned to triplets and the others are assigned to singlets
or doublets, then $W_B$ is forbidden. This provides a solution to the proton life time problem.

\subsection{Exotic Quark Decay and Proton Decay Suppression}

The absence of $W_B$ makes exotic quarks and proton stable, 
but the existence of exotic quarks which have life time longer than 0.1 second spoils the success of
Big Ban nucleosynthesis.
In order to evade this problem, the $S_4$ symmetry must be broken. 
Therefore, it is assumed that the $S_4$ breaking terms are induced from non-  renormalizable terms.
We introduce $G_2$ singlet $T$ as triplet of $S_4$ and add the quartic terms:
\eqn{
W_{NRB}=\frac{1}{M_P}T\L(QQg+g^cU^cD^c+gE^cU^c+g^cLQ+gD^cN^c\R).
}
Where the order one coefficients in front of each terms are omitted for simplicity.
When $T$ develops VEV with
\eqn{
\frac{\L<T\R>}{M_P}\sim 10^{-12},    \label{condition}
}
the phenomenological constraints on the life times of proton and exotic quarks are satisfied at the same time
\cite{f-extra-u1}. The violation of $S_4$ symmetry gives $S_4$ breaking corrections to effective superpotential 
through the non-renormalizable terms which are expressed in the same manner as Eq.(10):
\eqn{
W_{NRFV}=\frac{1}{M^2_P}T^2\L(H^UQU^c+H^DQD^c+H^DLE^c+ H^ULN^c+M'N^cN^c+SH^UH^D\R)+\frac{1}{M_P}TSgg^c.
}
 Since the above corrections are negligibly small, 
the $S_4$ flavor symmetry approximately holds in low energy effective theory.
One finds that the most economical flavon sector is the one which is exchanged $T$ into 
{\em superfield-product}; $\Phi\Phi^c/M_P$, by embedding $\Phi^c$ to a $S_4$ triplet
 (Hereafter, we call $\Phi$ and $\Phi^c$ as flavon which is the trigger of flavor violation.).  
 In this case, the condition 
of Eq. (\ref{condition}) correspond to the following relation:
\eqn{
\frac{\langle\Phi\rangle \langle\Phi^c\rangle }{M^2_P}\sim 10^{-12},
} 
and then the right-handed neutrino mass scale can be predicted as follows:
\eqn{M_R\sim \langle\Phi\rangle\sim 10^{-6}M_P\sim 10^{12}\ {\rm GeV}.}
Hence, by applying the above relation to the measurement of proton and 
exotic quarks (In our model, we call exotic quarks as $g$-quark.) lifetime, 
it is expected that one can determine the right-handed neutrino mass scale.

\section{The Model A}

Hereafter, we concentrate on $W_0$ that contributes mass matrices of quarks and leptons.
Although the $S_4$ symmetry reduces the Yukawa coupling constants,
there is still an overabundance of parameters.
In order to reduce the Yukawa coupling constants further, 
we extend the flavor symmetry to $S_4\times Z_2$ \cite{lepton}.
As a consequence, we can construct many predictive models.
But we do not give the perfect classification of such models, and we only give two
example models.
Firstly, we give the model A which gives the prediction to the lepton sector.
In model A, the quark, lepton, Higgs and flavon 
superfields are assigned to 
the representations of $S_4\times Z_2$ as Table 2 and Table 4. 

The superpotential $W_0$ which is consistent with $G_2$ and the symmetries of Table 2 and Table 4 is given by
\eqn{
W_0&=&Y^U_1H^U_3(Q_1U^c_1+Q_2U^c_2)+Y^U_3H^U_3Q_3U^c_3\no \\
&+&Y^U_4Q_3(H^U_1U^c_1+H^U_2U^c_2)+Y^U_5(H^U_1Q_1+H^U_2Q_2)U^c_3\no \\
&+&Y^D_1H^D_3(Q_1D^c_1+Q_2D^c_2)+Y^D_3H^D_3Q_3D^c_3\no \\
&+&Y^D_4Q_3(H^D_1D^c_1+H^D_2D^c_2)+Y^D_5(H^D_1Q_1+H^D_2Q_2)D^c_3\no \\
&+&Y^N_2\L[H^U_1(L_1N^c_2+L_2N^c_1)+H^U_2(L_1N^c_1-L_2N^c_2)\R] \no \\
&+&Y^N_3H^U_3L_3N^c_3+Y^N_4L_3(H^U_1N^c_1+H^U_2N^c_2) \no \\
&+&Y^E_1E^c_1(H^D_1L_1+H^D_2L_2)+Y^E_2E^c_2H^D_3L_3+Y^E_3E^c_3(H^D_1L_2-H^D_2L_1) \no \\
&+&Y^M_1\Phi(N^c_1N^c_1+N^c_2N^c_2)+Y^M_3\Phi N^c_3N^c_3.
}
There are sixteen complex Yukawa coupling constants in this superpotential.
The twelve phases of these can be absorbed by 
redefinition of the five of six quark superfields
$\{Q_i,Q_3,U^c_i,U^c_3,D^c_i,D^c_3\}$ and seven lepton superfields $\{L_i,L_3,E^c_1,E^c_2,E^c_3,N^c_i,N^c_3\}$.
Without loss of generality, we can define
$Y^U_{3,4,5},Y^D_{4,5},Y^N_{2,4},Y^E_{1,2,3},Y^M_{1,3}$ to be real.
We define the phases of complex Yukawa couplings as follows:
\eqn{
Y^U_1=e^{i\alpha}|Y^U_1|,\quad Y^D_1=e^{i\beta}|Y^D_1|,\quad Y^D_3=e^{i\gamma}|Y^D_3|,\quad Y^N_3=e^{i\delta}|Y^N_3|.
}
We write the VEV of the flavon 
as
\eqn{
\L<\Phi\R>=V,
}
and the VEVs of the $SU(2)_W$ doublet Higgses as
\eqn{
&&\L<H^U_1\R>=v_u\cos\theta_u,\quad \L<H^U_2\R>=v_u\sin\theta_u,\quad \L<H^U_3\R>=v'_u, \no \\
&&\L<H^D_1\R>=v_d\cos\theta_d,\quad \L<H^D_2\R>=v_d\sin\theta_d,\quad \L<H^D_3\R>=v'_d,
}
where we assume these VEVs are real and the parameters $V, v_{u,d}, v'_{u,d}$ are non-negative and the relation
\eqn{
\sqrt{v^2_u+v'^2_u+v^2_d+v'^2_d}=174\ {\rm GeV}
}
is satisfied. In this paper, we do not consider how these VEVs are derived from the Higgs potential.
If we define the non-negative mass parameters as follows:
\eqn{
\begin{tabular}{llll}
$M_1=Y^M_1V$,        & $M_3=Y^M_3V$,          &                     &   \\
$m^u_1=|Y^U_1|v'_u$, & $m^u_3=Y^U_3v'_u$,     & $m^u_4=Y^U_4v_u$,   & $m^u_5=Y^U_5v_u$,  \\
$m^d_1=|Y^D_1|v'_d$, & $m^d_3=|Y^D_3|v'_d$,   & $m^d_4=Y^D_4v_d$,   & $m^d_5=Y^D_5v_d$,  \\
$m^\nu_2=Y^N_2v_u$,  & $m^\nu_3=|Y^N_3|v'_u$, & $m^\nu_4=Y^N_4v_u$, &   \\
$m^l_1=Y^E_1v_d$,    & $m^l_2=Y^E_2v'_d$,     & $m^l_3=Y^E_3v_d$,   &
\end{tabular}
}
then the mass matrices of up-type quarks ($M_u$), down-type quarks ($M_d$), charged leptons ($M_l$),
Dirac neutrinos ($M_D$) and Majorana neutrinos ($M_R$) are given by
\eqn{
\begin{tabular}{ll}
$M_u=\Mat3{e^{i\alpha}m^u_1}{0}{m^u_5\cos\theta_u}
{0}{e^{i\alpha}m^u_1}{m^u_5\sin\theta_u}
{m^u_4\cos\theta_u}{m^u_4\sin\theta_u}{m^u_3}$, &
$M_d=\Mat3{e^{i\beta}m^d_1}{0}{m^d_5\cos\theta_d}
{0}{e^{i\beta}m^d_1}{m^d_5\sin\theta_d}
{m^d_4\cos\theta_d}{m^d_4\sin\theta_d}{e^{i\gamma}m^d_3}$,  \\
$M_l=\Mat3{m^l_1\cos\theta_d}{0}{-m^l_3\sin\theta_d}
{m^l_1\sin\theta_d}{0}{m^l_3\cos\theta_d}
{0}{m^l_2}{0}$, &
$M_D=\Mat3{m^\nu_2\sin\theta_u}{m^\nu_2\cos\theta_u}{0}
{m^\nu_2\cos\theta_u}{-m^\nu_2\sin\theta_u}{0}
{m^\nu_4\cos\theta_u}{m^\nu_4\sin\theta_u}{e^{i\delta}m^\nu_3}$,  \\
$M_R=\Mat3{M_1}{0}{0}{0}{M_1}{0}{0}{0}{M_3}$. &
\end{tabular}
}
After the seesaw mechanism, the light neutrino mass matrix is given by
\eqn{
M_\nu&=&M_DM^{-1}_RM^t_D=\Mat3{\rho^2_2}{0}{\rho_2\rho_4\sin2\theta_u}
{0}{\rho^2_2}{\rho_2\rho_4\cos2\theta_u}
{\rho_2\rho_4\sin2\theta_u}{\rho_2\rho_4\cos2\theta_u}{\rho^2_4+e^{2i\delta}\rho^2_3},
}
where
\eqn{
\rho_2=\frac{m^\nu_2}{\sqrt{M_1}},\quad \rho_4=\frac{m^\nu_4}{\sqrt{M_1}},\quad \rho_3=\frac{m^\nu_3}{\sqrt{M_3}}.
}
In the lepton sector, the mass eigenvalues and diagonalization matrix of charged leptons are given by
\eqn{
V^\dagger_l M^*_l M^t_l V_l&=&diag(m^2_e,m^2_\mu, m^2_\tau)=((m^l_2)^2,(m^l_3)^2,(m^l_1)^2), \\
V_l&=&\Mat3{0}{-\sin\theta_d}{\cos\theta_d}
{0}{\cos\theta_d}{\sin\theta_d}
{-1}{0}{0},
}
and those of the light neutrinos are given by
\eqn{
V^t_\nu M_\nu V_\nu&=&diag(e^{i(\phi_1-\phi)}m_{\nu_1},e^{i(\phi_2+\phi)}m_{\nu_2},m_{\nu_3}), \\
V_\nu&=&\Mat3{\sin2\theta_u}{-\cos2\theta_u}{0}
{\cos2\theta_u}{\sin2\theta_u}{0}
{0}{0}{1}
\Mat3{-\sin\theta_\nu}{e^{i\phi}\cos\theta_\nu}{0}
{0}{0}{1}
{e^{-i\phi}\cos\theta_\nu}{\sin\theta_\nu}{0},
}
from Eq.(25) and Eq.(27), the Maki-Nakagawa-Sakata (MNS) matrix is given by
\eqn{
V'_{MNS}&=&V^\dagger_lV_\nu P_\nu=\Mat3{-e^{-i\phi}\cos\theta_\nu}{-\sin\theta_\nu}{0}
{-\cos\bar{\theta}\sin\theta_\nu}{e^{i\phi}\cos\bar{\theta}\cos\theta_\nu}{\sin\bar{\theta}}
{-\sin\bar{\theta}\sin\theta_\nu}{e^{i\phi}\sin\bar{\theta}\cos\theta_\nu}{-\cos\bar{\theta}}P_\nu,
}
where
\eqn{
&&\bar{\theta}=\theta_d+2\theta_u, \\
&&P_\nu=diag(e^{-i(\phi_1-\phi)/2},e^{-i(\phi_2+\phi)/2},1).
}
Following ref. \cite{lepton}, we get
\eqn{
\tan^2\theta_\nu&=&\frac{\sqrt{m^2_{\nu_2}-m^2_{\nu_3}\sin^2\phi}-m_{\nu_3}|\cos\phi|}
{\sqrt{m^2_{\nu_1}-m^2_{\nu_3}\sin^2\phi}+m_{\nu_3}|\cos\phi|}, \\
\sin(\phi_1-\phi_2)&=&\frac{m_{\nu_3}\sin\phi}{m_{\nu_1}m_{\nu_2}}
\L[\sqrt{m^2_{\nu_2}-m^2_{\nu_3}\sin^2\phi}+\sqrt{m^2_{\nu_1}-m^2_{\nu_3}\sin^2\phi}\R], \\
\sin(\phi_1-\phi)&=&\frac{\sin\phi}{m_{\nu_1}}
\L[m_{\nu_3}\sqrt{1-\sin^2\phi}+\sqrt{m^2_{\nu_1}-m^2_{\nu_3}\sin^2\phi}\R].
}
After the redefinition of the fields, the MNS matrix is transformed to the standard form in Eq.(106)
where the parameters are given by
\eqn{
\theta_{13}=0,\quad \theta_{12}=\theta_\nu,\quad \theta_{23}=\bar{\theta},\quad
\alpha'=\frac{\phi_1-\phi_2}{2},\quad \beta'=\frac{\phi_1-\phi}{2}.
}
If the neutrino masses have been measured, the two Majorana phases
$\alpha'$ and $\beta'$ would be predicted by Eqs.(31), (32), (33) and (34).
In addition, $\theta_{13}=0$ is predicted, so totally three predictions are given in the lepton sector.

In the quark sector, the mass eigenvalues and diagonalization matrices of quarks are given as follows:
\eqn{
V^\dagger_uM^*_uM^t_uV_u&=&diag(m^2_u,m^2_c,m^2_t),\\
V_u&=&\Mat3{\cos\theta_u}{-\sin\theta_u}{0}{\sin\theta_u}{\cos\theta_u}{0}{0}{0}{e^{i\phi_u}}
\Mat3{\cos\theta'_u}{0}{\sin\theta'_u}{0}{1}{0}{-\sin\theta'_u}{0}{\cos\theta'_u}, \\
m^2_u&=&\frac12\L[(m^u_1)^2+(m^u_3)^2+(m^u_4)^2+(m^u_5)^2-\mu^2_u\R], \\
m^2_c&=&(m^u_1)^2, \\
m^2_t&=&\frac12\L[(m^u_1)^2+(m^u_3)^2+(m^u_4)^2+(m^u_5)^2+\mu^2_u\R], \\
\mu^2_u&=&\sqrt{\L((m^u_3)^2+(m^u_4)^2-(m^u_1)^2-(m^u_5)^2\R)^2+4R^2_u}, \\
R_u&=&\sqrt{(m^u_1m^u_4\cos\alpha+m^u_3m^u_5)^2+(m^u_1m^u_4\sin\alpha)^2}, \\
\tan2\theta'_u&=&\frac{2R_u}{(m^u_3)^2+(m^u_4)^2-(m^u_1)^2-(m^u_5)^2},\\
\tan\phi_u&=&\frac{m^u_1m^u_4\sin\alpha}{m^u_1m^u_4\cos\alpha+m^u_3m^u_5},\\
V^\dagger_dM^*_dM^t_dV_d&=&diag(m^2_d,m^2_s,m^2_b),\\
V_d&=&\Mat3{\cos\theta_d}{-\sin\theta_d}{0}{\sin\theta_d}{\cos\theta_d}{0}{0}{0}{e^{i\phi_d}}
\Mat3{\cos\theta'_d}{0}{\sin\theta'_d}{0}{1}{0}{-\sin\theta'_d}{0}{\cos\theta'_d}, \\
m^2_d&=&\frac12\L[(m^d_1)^2+(m^d_3)^2+(m^d_4)^2+(m^d_5)^2-\mu^2_d\R], \\
m^2_s&=&(m^d_1)^2, \\
m^2_b&=&\frac12\L[(m^d_1)^2+(m^d_3)^2+(m^d_4)^2+(m^d_5)^2+\mu^2_d\R], \\
\mu^2_d&=&\sqrt{\L((m^d_3)^2+(m^d_4)^2-(m^d_1)^2-(m^d_5)^2\R)^2+4R^2_d}, \\
R_d&=&\sqrt{(m^d_1m^d_4\cos\beta+m^d_3m^d_5\cos\gamma)^2+(m^d_1m^d_4\sin\beta-m^d_3m^d_5\sin\gamma)^2}, \\
\tan2\theta'_d&=&\frac{2R_d}{(m^d_3)^2+(m^d_4)^2-(m^d_1)^2-(m^d_5)^2}, \\
\tan\phi_d&=&\frac{m^d_1m^d_4\sin\beta-m^d_3m^d_5\sin\gamma}{m^d_1m^d_4\cos\beta+m^d_3m^d_5\cos\gamma},
}
from which the Cabbibo-Kobayashi-Maskawa (CKM) matrix is given by
\eqn{
&&V_{CKM}=V^\dagger_u V_d= \no \\
&&\Mat3{\cos\tilde{\theta}\cos\theta'_u\cos\theta'_d+e^{i\bar{\phi}}\sin\theta'_u\sin\theta'_d}
{-\sin\tilde{\theta}\cos\theta'_u}
{\cos\tilde{\theta}\cos\theta'_u\sin\theta'_d-e^{i\bar{\phi}}\sin\theta'_u\cos\theta'_d}
{\sin\tilde{\theta}\cos\theta'_d}{\cos\tilde{\theta}}{\sin\tilde{\theta}\sin\theta'_d}
{\cos\tilde{\theta}\sin\theta'_u\cos\theta'_d-e^{i\bar{\phi}}\cos\theta'_u\sin\theta'_d}
{-\sin\tilde{\theta}\sin\theta'_u}
{\cos\tilde{\theta}\sin\theta'_u\sin\theta'_d+e^{i\bar{\phi}}\cos\theta'_u\cos\theta'_d},
}
where
\eqn{
\tilde{\theta}=\theta_d-\theta_u,\quad \bar{\phi}=\phi_d-\phi_u. 
}
The experimental values of the matrix elements and Jarlskog invariant in Eq.(105) are reproduced by putting
\eqn{
\tilde{\theta}=13.3^\circ,\quad \theta'_u=10.2^\circ,\quad \theta'_d=10.4^\circ,\quad \bar{\phi}=1.1^\circ.
}
In ref. \cite{lepton}, it is assumed that the VEVs of Higgs $S_3$ doublets
are fixed in the direction of $\theta_u=\theta_d=\frac{\pi}{4}$, which enforces $\tilde{\theta}=0$
(and predicts the atmospheric neutrino mixing angle is maximal). 
This means the Cabbibo angle is zero.
In contrast, there is no such a condition of vacuum directions in the model A.

Due to an overabundance of free parameters, there is no prediction in quark sector.
But we can show that there exist consistent parameter sets.
For example, if we put
\eqn{
\begin{tabular}{llll}
$\alpha=3.08^\circ$, & $\beta=1.22^\circ$, & $\gamma=-1.10^\circ$,  &  \\
$m^u_1=624\ {\rm MeV}$,      & $m^u_3=170\ {\rm GeV}$,     & $m^u_4=3.47\ {\rm GeV}$,       & $m^u_5=30.5\ {\rm GeV}$,  \\
$m^d_1=55.0\ {\rm MeV}$,     & $m^d_3=2.84\ {\rm GeV}$,    & $m^d_4=300\ {\rm MeV}$,        & $m^d_5=522\ {\rm MeV}$,  \\
$m^l_1=1.75\ {\rm GeV}$,     & $m^l_2=487\ {\rm KeV}$,     & $m^l_3=103\ {\rm MeV}$,        &
\end{tabular}
}
then the quark masses in Eq.(104) and the parameters of CKM matrix in Eq.(55) are reproduced.
These parameters can be expressed by the perturbative Yukawa coupling constants and the VEVs of Higgs fields
through Eq.(20), for example as follows:
\eqn{
\begin{tabular}{llll}
$v_u=32\ {\rm GeV}$,                     & $v'_u=170\ {\rm GeV}$,                  
 & $v_d=9.9\ {\rm GeV}$,        & $v'_d=14.2\ {\rm GeV}$,  \\
$\L|Y^U_1\R|=3.7\times 10^{-3}$, & $\L|Y^U_3\R|=1.0$,               & $\L|Y^U_4\R|=0.11$,  & $\L|Y^U_5\R|=0.95$, \\
$\L|Y^D_1\R|=3.9\times 10^{-3}$, & $\L|Y^D_3\R|=0.20$,              & $\L|Y^D_4\R|=0.030$, & $\L|Y^D_5\R|=0.050$,  \\
$\L|Y^E_1\R|=0.18$,              & $\L|Y^E_2\R|=3.4\times 10^{-5}$, & $\L|Y^E_3\R|=0.010$. & 
\end{tabular}
}
As all the coupling constants of the model are perturbative,
it is consistent that the fundamental energy scale is much larger than the electroweak scale,
which is the base of naturalness problem.
It is noted that there may be the problem of flavor-changing neutral currents
which are generally enhanced in multi-Higgs models thus ruling out such models \cite{higgs}.
But this is beyond the scope of this paper, we leave this subject for a future work.


\section{The Model B}

Next, we give the model B which gives the prediction to the quark sector.
In the model B, the quark, lepton, Higgs and flavon 
superfields are assigned to 
the representations of $S_4\times Z_2$ as Table 3 and Table 4. 

The superpotential $W_0$ which is consistent with $G_2$ and the symmetries of Table 3 and Table 4 is given by
\eqn{
W_0&=&Y^U_2\L[H^U_1(Q_1U^c_2+Q_2U^c_1)+H^U_2(Q_1U^c_1-Q_2U^c_2)\R] \no \\
&+&Y^U_4Q_3(H^U_1U^c_1+H^U_2U^c_2)+Y^U_5(H^U_1Q_1+H^U_2Q_2)U^c_3 \no \\
&+&Y^D_2\L[H^D_1(Q_1D^c_2+Q_2D^c_1)+H^D_2(Q_1D^c_1-Q_2D^c_2)\R] \no \\
&+&Y^D_4Q_3(H^D_1D^c_1+H^D_2D^c_2)+Y^D_5(H^D_1Q_1+H^D_2Q_2)D^c_3 \no \\
&+&Y^N_2\L[H^U_1(L_1N^c_2+L_2N^c_1)+H^U_2(L_1N^c_1-L_2N^c_2)\R] \no \\
&+&Y^N_3L_3H^U_3N^c_3+Y^N_5(H^U_1L_1+H^U_2L_2)N^c_3 \no \\
&+&Y^E_1H^D_3(L_1E^c_1+L_2E^c_2)+Y^E_3L_3E^c_3H^D_3\no \\
&+&Y^E_4L_3(H^D_1E^c_1+H^D_2E^c_2)+Y^E_5(H^D_1L_1+H^D_2L_2)E^c_3 \no \\
&+&Y^M_2\L[2\Phi_1N^c_1N^c_2+\Phi_2(N^c_1N^c_1-N^c_2N^c_2)\R]+2Y^M_4(\Phi_1N^c_1N^c_3+\Phi_2N^c_2N^c_3).
}
As implemented in previous section, we define the Yukawa couplings
$Y^U_{2,4,5},Y^D_{4,5},Y^N_{2,3},Y^E_{4,5},Y^M_{2,4}$ to be real 
and define the phases of the complex Yukawa couplings as follows:
\eqn{
Y^D_2=e^{i\delta}|Y^D_2|,\quad Y^N_5=e^{i\alpha}|Y^N_5|,\quad Y^E_1=e^{i\beta}|Y^E_1|,\quad Y^E_3=e^{i\gamma}|Y^E_3|.
}
The definitions of the VEVs of flavon 
scalar fields are replaced from Eq.(17) to
\eqn{
\L<\Phi_1\R>=V\cos\theta,\quad \L<\Phi_2\R>=V\sin\theta.
}
But those of the VEVs of Higgs fields are the same as Eq.(18).
 In this case, the flavor symmetry is explicitly broken by the threshold correction due  to the mass differences of RHNs. But here, we assume that the correction is negligible.
The non-negative mass parameters are given as follows:
\eqn{
\begin{tabular}{llll}
$M_2=Y^M_2V$,        & $M_4=Y^M_4V$,         &                      & \\
$m^u_2=Y^U_2v_u$,    & $m^u_4=Y^U_4v_u$,     & $m^u_5=Y^U_5v_u$,     & \\
$m^d_2=|Y^D_2|v_d$,  & $m^d_4=Y^D_4v_d$,     & $m^d_5=Y^D_5v_d$,     & \\
$m^\nu_2=Y^N_2v_u$,  & $m^\nu_3=Y^N_3v'_u$,  & $m^\nu_5=|Y^N_5|v_u$, & \\
$m^l_1=|Y^E_1|v'_d$, & $m^l_3=|Y^E_3|v'_d$,  & $m^l_4=Y^E_4v_d$,     & $m^l_5=Y^E_5v_d$,
\end{tabular}
}
then the mass matrices of quarks and leptons are given by
\eqn{
\begin{tabular}{ll}
$M_u=\Mat3{m^u_2\sin\theta_u}{m^u_2\cos\theta_u}{m^u_5\cos\theta_u}
{m^u_2\cos\theta_u}{-m^u_2\sin\theta_u}{m^u_5\sin\theta_u}
{m^u_4\cos\theta_u}{m^u_4\sin\theta_u}{0}$, &
$M_d=\Mat3{e^{i\delta}m^d_2\sin\theta_d}{e^{i\delta}m^d_2\cos\theta_d}{m^d_5\cos\theta_d}
{e^{i\delta}m^d_2\cos\theta_d}{-e^{i\delta}m^d_2\sin\theta_d}{m^d_5\sin\theta_d}
{m^d_4\cos\theta_d}{m^d_4\sin\theta_d}{0}$, \\
$M_l=\Mat3{e^{i\beta}m^l_1}{0}{m^l_5\cos\theta_d}
{0}{e^{i\beta}m^l_1}{m^l_5\sin\theta_d}
{m^l_4\cos\theta_d}{m^l_4\sin\theta_d}{e^{i\gamma}m^l_3}$, &
$M_D=\Mat3{m^\nu_2\sin\theta_u}{m^\nu_2\cos\theta_u}{e^{i\alpha}m^\nu_5\cos\theta_u}
{m^\nu_2\cos\theta_u}{-m^\nu_2\sin\theta_u}{e^{i\alpha}m^\nu_5\sin\theta_u}
{0}{0}{m^\nu_3}$, \\
$M_R=\Mat3{M_2\sin\theta}{M_2\cos\theta}{M_4\cos\theta}
{M_2\cos\theta}{-M_2\sin\theta}{M_4\sin\theta}
{M_4\cos\theta}{M_4\sin\theta}{0}$. &
\end{tabular}
}
Here we assume that the mass parameters are hierarchical as $m^u_5\gg m^u_2 \gg m^u_4$ and
$m^d_5\gg m^d_2 \gg m^d_4$, then the mass eigenvalues and diagonalization matrices of quarks are 
approximately given as follows:
\eqn{
V^\dagger_uM^*_uM^t_uV_u&=&diag(m^2_u,m^2_c,m^2_t),  \\
V_u&\simeq&\Mat3{\sin\theta_u}{\cos\theta_u}{0}
{-\cos\theta_u}{\sin\theta_u}{0}
{0}{0}{1}
\Mat3{\cos\theta'_u}{-\sin\theta'_u}{0}
{0}{0}{1}
{-\sin\theta'_u}{-\cos\theta'_u}{0},  \\
m^2_u&\simeq&\L(m^u_4\sin3\theta_u\R)^2,  \\
m^2_c&\simeq&(m^u_2)^2+(m^u_4\cos3\theta_u)^2,  \\
m^2_t&\simeq&(m^u_5)^2,  \\
\theta'_u&\simeq&\frac{\pi}{2}+\frac{m^u_2m^u_4\cos3\theta_u}{(m^u_2)^2-(m^u_4)^2},  \\
V^\dagger_dM^*_dM^t_dV_d&=&diag(m^2_d,m^2_s,m^2_b),  \\
V_d&\simeq&\Mat3{\sin\theta_d}{\cos\theta_d}{0}
{-\cos\theta_d}{\sin\theta_d}{0}
{0}{0}{e^{i\delta}}
\Mat3{\cos\theta'_d}{-\sin\theta'_d}{0}
{0}{0}{1}
{-\sin\theta'_d}{-\cos\theta'_d}{0},  \\
m^2_d&\simeq&\L(m^d_4\sin3\theta_d\R)^2,  \\
m^2_s&\simeq&(m^d_2)^2+(m^d_4\cos3\theta_d)^2,  \\
m^2_b&\simeq&(m^d_5)^2, \no \\
\theta'_d&\simeq&\frac{\pi}{2}+\frac{m^d_2m^d_4\cos3\theta_d}{(m^d_2)^2-(m^d_4)^2},
}
from these the CKM matrix is given by
\eqn{
&&V_{CKM}=V^\dagger_uV_d= \no \\
&&\Mat3{\cos\bar{\theta}\cos\theta'_u\cos\theta'_d+e^{i\delta}\sin\theta'_u\sin\theta'_d}
{-\cos\bar{\theta}\cos\theta'_u\sin\theta'_d+e^{i\delta}\sin\theta'_u\cos\theta'_d}
{-\sin\bar{\theta}\cos\theta'_u}
{-\cos\bar{\theta}\sin\theta'_u\cos\theta'_d+e^{i\delta}\cos\theta'_u\sin\theta'_d}
{\cos\bar{\theta}\sin\theta'_u\sin\theta'_d+e^{i\delta}\cos\theta'_u\cos\theta'_d}
{\sin\bar{\theta}\sin\theta'_u}
{\sin\bar{\theta}\cos\theta'_d}{-\sin\bar{\theta}\sin\theta'_d}{\cos\bar{\theta}}, \no \\
&&\bar{\theta}=\theta_d-\theta_u.
}
As the ten observables which consist of six mass eigenvalues, three CKM mixing angles and one phase
are described by 9 parameters $m^u_{2,4,5},m^d_{2,4,5},\theta_{u,d},\delta$,
there is one prediction in quark sector. If we put
\eqn{
\begin{tabular}{lll}
$m^u_4=55.0\ {\rm MeV}$,       &  $m^u_2=550\ {\rm MeV}$,        & $m^u_5=172.5\ {\rm GeV}$,  \\
$m^d_4=13.5\ {\rm MeV}$,       &  $m^d_2= 65.0\ {\rm MeV}$,      & $m^d_5=2.89\ {\rm GeV}$,   \\
$\theta_u=0.60^\circ$, &  $\theta_d=2.89^\circ$, & $\delta=92.37^\circ$,
\end{tabular}
}
then the observables are given by
\eqn{
&&
\begin{tabular}{lll}
$|V_{ud}|=0.974$,   & $|V_{us}|=0.226$,  & $|V_{ub}|=0.00398$, \\
$|V_{cd}|=0.226$,   & $|V_{cs}|=0.973$,  & $|V_{cb}|=0.0398$, \\
$|V_{td}|=0.00804$, & $|V_{ts}|=0.0391$, & $|V_{tb}|=0.999$, 
\end{tabular} \no \\
&&J_{CP}=3.18\times 10^{-5}, \no \\
&&
\begin{tabular}{lll}
$m_t=172.5\ {\rm GeV}$, & $m_c=553\ {\rm MeV}$,  & $m_u=1.73\ {\rm MeV}$,  \\
$m_b=2.89\ {\rm GeV}$,  & $m_s=66.4\ {\rm MeV}$, & $m_d=2.04\ {\rm MeV}$, 
\end{tabular}
}
which are marginally consistent with Eqs.(104) and (105).

In the lepton sector, we just show an example of the set of parameters below,
since there are too many free parameters to give predictions.
For simplicity, we assume $\alpha=0, m^\nu_3=0$.
Then, the seesaw neutrino mass matrix is given by
\eqn{
M_\nu&=&M_DM^{-1}_RM^t_D=\rho\Mat3{A}{B}{0}{B}{C}{0}{0}{0}{0},
}
where
\eqn{
&&A=\cos^2(\theta+\theta_u)-2r\cos(2\theta-\theta_u)\cos\theta_u+r^2\cos^2\theta_u, \no \\
&&B=-\cos(\theta+\theta_u)\sin(\theta+\theta_u)-r\sin2\theta+r^2\cos\theta_u\sin\theta_u, \no \\
&&C=\sin^2(\theta+\theta_u)-2r\sin(2\theta-\theta_u)\sin\theta_u+r^2\sin^2\theta_u, \no \\
&&r=\frac{M_2m^\nu_5}{M_4m^\nu_2},\quad \rho=-\frac{(m^\nu_2)^2}{M_2\sin3\theta}.
}
The mass eigenvalues and diagonalization matrix of charged leptons are given by
\eqn{
V^\dagger_lM^*_lM^t_lV_l&=&diag(m^2_e,m^2_\mu,m^2_\tau),  \\
V_l&=&\Mat3{\cos\theta_d}{-\sin\theta_d}{0}{\sin\theta_d}{\cos\theta_d}{0}{0}{0}{e^{i\phi_l}}
\Mat3{0}{\cos\theta_l}{\sin\theta_l}
{-1}{0}{0}
{0}{-\sin\theta_l}{\cos\theta_l},  \\
m^2_e&=&(m^l_1)^2,  \\
m^2_\mu&=&\frac12\L[(m^l_1)^2+(m^l_3)^2+(m^l_4)^2+(m^l_5)^2-\mu^2_l\R], \\
m^2_\tau&=&\frac12\L[(m^l_1)^2+(m^l_3)^2+(m^l_4)^2+(m^l_5)^2+\mu^2_l\R], \\
\mu^2_l&=&\sqrt{\L((m^l_3)^2+(m^l_4)^2-(m^l_1)^2-(m^l_5)^2\R)^2+4R^2_l}, \\
R_l&=&\sqrt{(m^l_1m^l_4\cos\beta+m^l_3m^l_5\cos\gamma)^2+(m^l_1m^l_4\sin\beta-m^l_3m^l_5\sin\gamma)^2}, \\
\tan2\theta_l&=&\frac{2R_l}{(m^l_3)^2+(m^l_4)^2-(m^l_1)^2-(m^l_5)^2}, \\
\tan\phi_l&=&\frac{m^l_1m^l_4\sin\beta-m^l_3m^l_5\sin\gamma}{m^l_1m^l_4\cos\beta+m^l_3m^l_5\cos\gamma},
}
and those of the light neutrinos are given by
\eqn{
V^t_\nu M_\nu V_\nu&=&diag(m_{\nu_1},m_{\nu_2},m_{\nu_3}),  \\
V_\nu&=&\Mat3{\cos\theta_\nu}{\sin\theta_\nu}{0}{-\sin\theta_\nu}{\cos\theta_\nu}{0}{0}{0}{1},  \\
\tan2\theta_\nu&=&\frac{2B}{C-A},  \\
m_{\nu_1}&=&\frac12\rho\L[A+C-\sqrt{(A-C)^2+4B^2}\R],  \\
m_{\nu_2}&=&\frac12\rho\L[A+C+\sqrt{(A-C)^2+4B^2}\R],  \\
m_{\nu_3}&=&0.
}
From the above equations, the MNS matrix is given by
\eqn{
V_{MNS}&=&V^\dagger_lV_\nu=\Mat3{\cos\theta_{12}}{\sin\theta_{12}}{0}
{-\sin\theta_{12}\cos\theta_l}{\cos\theta_{12}\cos\theta_l}{-e^{-i\phi_l}\sin\theta_l}
{-\sin\theta_{12}\sin\theta_l}{\cos\theta_{12}\sin\theta_l}{e^{-i\phi_l}\cos\theta_l}, \no \\
\theta_{12}&=&\theta_d+\theta_\nu-\frac{\pi}{2}.
}
By the redefinition of fields, the phase $\phi_l$ is rotated away. If we put 
\eqn{
&&r=0.417,\quad \theta=89.39^\circ,\no \\
&&m^l_1=487\ {\rm KeV},\quad m^l_3=1226\ {\rm MeV},\quad m^l_4=148.8\ {\rm MeV},\quad m^l_5=1235\ {\rm MeV},
}
then the charged lepton masses in Eq.(104) are reproduced
and the mixing angles and the ratio of two neutrino squared mass differences are given by
\eqn{
\theta_{13}=0^\circ,\quad 
\theta_{12}=33.93^\circ,\quad \theta_{23}=45^\circ,\quad \frac{\Delta m^2_{21}}{\Delta m^2_{23}}=0.032,
}
which are consistent with Eq.(106).
The parameters in Eq.(75) and (95) can be well explained 
by the perturbative Yukawa coupling constants and the VEVs of Higgs fields
through Eq.(61); for example, as follows:
\eqn{
\begin{tabular}{llll}
$v_u=172.5\ {\rm GeV}$,             & $v'_u=22.8\ {\rm GeV}$,             & $v_d=14.5\ {\rm GeV}$,   &$v'_d=9.4\ {\rm GeV}$, \\
$|Y^U_2|=3.2\times 10^{-3}$,& $|Y^U_4|=3.2\times 10^{-4}$,& $|Y^U_5|=1.0$,   &  \\
$|Y^D_2|=4.5\times 10^{-3}$,& $|Y^D_4|=9.3\times 10^{-4}$,& $|Y^D_5|=0.20$,  &  \\
$|Y^E_1|=5.2\times 10^{-5}$,& $|Y^E_3|=0.13$,             & $|Y^E_4|=0.010$, & $|Y^E_5|=0.085$. 
\end{tabular}
}

\section{Flavor Violation in Extended Sector}

In this chapter, we show a outline how to realize the flavor symmetry breaking 
in our model. 

\subsection{ Higgs and $g$-quark Sector }
 
At first, we discuss the flavor symmetry breaking in low energy scale.
In the both models; A and B, since the each of the representations for $S$, $H^U$, $H^D$, and $g(g^c)$ 
is chosen to the same assignments, $G_2$ and $S_4\times Z_2$ invariant superpotential; $W_S$, are given by

\eqn{
W_S &=& \lambda_1S_3 (H^U_1H^D_1 + H^U_2H^D_2) + \lambda_3S_3 H^U_3H^D_3\no\\
&+&  \lambda_4 H^U_3 (S_1H^D_1 + S_2H^D_2) +   \lambda_5 H^D_3 (S_1H^U_1 + S_2H^U_2)\no\\
&+& k S_3 (g_1g^c_1+g_2g^c_2 + g_3g^c_3),
}   
where one can take, without any loss of the generalities, 
$\lambda_{1,3,4,5}$ and $k$ as real, by redefinining an arbitrary field of $\{g,g^c\}$ 
and four arbitrary fields of $\{S_i, S_3, H^U_i, H^U_3, H^D_i, H^D_3\}$, respectively.
However, this superpotential could have would-be goldstone bosons when all of the Higgs fields acquire VEVs,   
because of an accidental $O(2)$ symmetry 
induced by the common rotation of the $S_4$ doublet.   
In order to avoid the problem, we assume that the flavor symmetry should be explicitly broken in the soft scalar mass terms, which can play role in giving the controllable parameters for the direction of the $SU(2)$ doublet Higgs VEVs.
Moreover, if these soft mass parameters are assumed to be real, 
parameters appeared in the Higgs potential can be also taken to be real. 
Hence these VEVs can be expected to be real. But the serious analysis for the Higgs potential is beyond our paper.

Here, we must discuss some potential problems; the Higgs mediated flavor changing neutral currents (FCNCs), and SUSY-FCNCs.
The first problem comes from multiple Higgs interactions with leptons and quarks.
The second one arises from the SUSY breaking terms, in which the flavor symmetry breaking is also introduced.

To resolve these problems, we assume that the soft breaking scale; $M_{SB}$, should be large enough to be satisfied  with $M_{SB}  > O(10 ~{\rm TeV})$ \cite{higgs}.

As a result of the gauge symmetry breaking, one finds the following degenerated mass matrix with diagonal form 
in the $g$-quark sector: 
 \eqn{
M_g=diag(m_g,m_g,m_g), \quad m_g = k \langle S_3\rangle.
} 
Note that such a degenerated mass is obtained only in case of coupling to the $S_3$ field. If the $Z_2$ assignments for
$g$ and $g^c$ fields are chosen to be opposite each other, then the last term of the Eq. (98) is modified as follows:
\eqn{
k[  \sqrt3 S_1 (g_2g^c_ 2 - g_3g^c_3 ) + S_2 (g_2g^c_ 2 + g_3g^c_3 -2 g_1g^c_ 1  )   ].
}
In this case, one straightforwardly finds that the non-degenerated diagonal mass matrix is obtained. However note that the mass matrix for the $g(g^c)$-superpartner depends on the form of the flavor symmetry breaking in the soft breaking mass term.

\subsection {Flavon Sector}

Next, we discuss the flavon sector which leads to the flavor symmetry breaking in high energy scale.

Considering the fact that there are not any renormalizable superpotentials for $\Phi$ and $\Phi^c$ because of the flavor and gauge symmetry, the lowest superpotential, (which is not renormalizable), is given as
\eqn{
W_{\Phi}
&=&
\frac{a_1}{2 M_P} \Phi^2_3[  (\Phi^c_1)^2 + (\Phi^c_2)^2 + (\Phi^c_3)^2 ]\nn\\
&+&
\frac{a_2}{2 M_P} (\Phi^2_1+ \Phi^2_2)[  (\Phi^c_1)^2 + (\Phi^c_2)^2 + (\Phi^c_3)^2 ]\nn\\
&+&
\frac{a_3}{2 M_P} \{ 2\sqrt3\Phi_1\Phi_2[  (\Phi^c_2)^2 - (\Phi^c_3)^2] + (\Phi^2_1- \Phi^2_2)[ (\Phi^c_2)^2 + (\Phi^c_3)^2 - 2(\Phi^c_1)^2 ]  \}.
}
In both of the models, one can expect that the soft breaking mass terms for $\Phi_3$( for model $A$) and $\Phi_i$( for model $B$) induce the $U(1)_Z$ symmetry breaking, by driven the soft breaking mass terms to be minus by the Yukawa coupling $Y^M \Phi N^c N^c$. As a subsequent result,  the following relation is found by solving the minimal condition for the potential;
\eqn{
\langle\Phi\rangle \sim \langle \Phi^c\rangle \sim \left( \frac{M_{SB} M_P}{a_i}  \right)^{1/2}
\sim 10^{11} {\rm GeV}a^{-1/2}_i  \left( \frac{M_{SB}}{10\ {\rm TeV} }  \right)^{1/2},
}
where the condition of Eq. (\ref{condition}) are satisfied in the case of 
\eqn{
a_i \sim 10^{-2}.
}
In a similar way of the low energy case, the direction of the VEVs are controlled by the 
explicit flavor symmetry breaking terms with the soft breaking.


\section{Summary}

In this paper, we have considered the $S_4\times Z_2$ flavor symmetric
extra U(1) model, and have obtained the following results:
\begin{enumerate}
\item  By assigning 
that exotic quarks are in $S_4$ triplets and the others are in singlets or doublets,
well-suppression of proton decay is realized.
\item The phenomenological constraints for the proton decay and the $g$-quark life time lead to a prediction that 
the right-handed neutrino mass should be around $10^{12}\ {\rm GeV}$.
\item The flavor symmetry leads to reduced parameters 
enough to give predictions to the lepton sector (in model A) or the quark sector (in model B).
\end{enumerate}
It might be expected that the new gauge symmetry 
 and the flavor symmetry are tested in LHC or future colliders.

\section*{Acknowledgments}
H.O. thanks great hospitality of the Kanazawa Institute for Theoretical
Physics at Kanazawa University. Discussions during my visit  were useful to
finalize this project.
This work is partially
supported by the ICTP grant Proj-30 and the Egyptian Academy for
Scientific Research and Technology (H.O.).

\appendix

\section*{Appendix}

\section{Experimental Values}

Running masses of quarks and charged leptons \cite{mass}:
\eqn{
\begin{tabular}{lll}
$m_u(m_Z)=1.28^{+0.50}_{-0.39} ({\rm MeV})$, &
$m_c(m_Z)=624\pm 83\ ({\rm MeV})$, &
$m_t(m_Z)=172.5\pm 3.0 ({\rm GeV})$, \\
$m_d(m_Z)=2.91^{+1.24}_{-1.20}\ ({\rm MeV})$, &
$m_s(m_Z)=55^{+16}_{-15}\ ({\rm MeV})$, &
$m_b(m_Z)=2.89\pm 0.09\ ({\rm GeV})$, \\
$m_e(m_Z)=0.48657\ ({\rm MeV})$, &
$m_\mu(m_Z)=102.72\ ({\rm MeV})$, &
$m_\tau(m_Z)=1746\ ({\rm MeV})$.
\end{tabular}
}
CKM matrix elements and Jarlskog invariant \cite{PDG2008}:
\eqn{
&&
\begin{tabular}{lll}
$\L| V_{ud}\R|=0.97418\pm0.00027$, &
$\L| V_{us}\R|=0.2255\pm0.0019$,   &
$\L| V_{ub}\R|=(3.93\pm0.36)\times 10^{-3}$, \\
$\L| V_{cd}\R|=0.230\pm0.011$, &
$\L| V_{cs}\R|=1.04\pm0.06$,   &
$\L| V_{cb}\R|=(41.2\pm1.1)\times 10^{-3}$, \\
$\L| V_{td}\R|=(8.1\pm0.6)\times10^{-3}$, &
$\L| V_{ts}\R|=(38.7\pm2.3)\times10^{-3}$,&
$\L| V_{tb}\R|>0.74$,
\end{tabular} \no \\
&&J_{CP}=Im(V_{ud}V^*_{ub}V_{cd}V^*_{cb})=(3.05^{+0.19}_{-0.20})\times10^{-5}.
}
Neutrino mass-squared differences and the parameters of MNS matrix \cite{PDG2008}:
\begin{eqnarray}
&&\Delta m^2_{21}= m^2_{\nu_2}-m^2_{\nu_1}=(8.0\pm 0.3)\times 10^{-5} \quad (eV^2), \quad
\Delta m^2_{32} = \L|m^2_{\nu_3}-m^2_{\nu_2}\R|=(1.9-3.0)\times 10^{-3}\quad (eV^2), \no \\
&&V_{MNS}=\Mat3{c_{12}c_{13}}{s_{12}c_{13}}{s_{13}e^{-i\delta'}}
{-s_{12}c_{23}-c_{12}s_{23}s_{13}e^{i\delta'}}{c_{12}c_{23}-s_{12}s_{23}s_{13}e^{i\delta'}}{s_{23}c_{13}}
{s_{12}s_{23}-c_{12}c_{23}s_{13}e^{i\delta'}}{-c_{12}s_{23}-s_{12}c_{23}s_{13}e^{i\delta'}}{c_{23}c_{13}}
\Mat3{1}{0}{0}{0}{e^{i\alpha'}}{0}{0}{0}{e^{i\beta'}}, \no \\
&&\theta_{12}=34.0^\circ {}^{+1.3^\circ}_{-1.5^\circ}, \quad
45.0^\circ > \theta_{23}>36.8^\circ, \quad
12.9^\circ >\theta_{13}>0.0^\circ.
\end{eqnarray}
\newpage

\section{Representations of Superfields}

\begin{table}[htbp]
\begin{center}
\begin{tabular}{|c|c|c|c|c|c|c|c|c|c|c|c||c|c|}
\hline
         &$Q$ &$U^c$    &$E^c$&$D^c$    &$L$ &$N^c$&$H^D$&$g^c$    &$H^U$&$g$ &$S$ &$\Phi$&$\Phi^c$\\ \hline
$SU(3)_c$&$3$ &$3^*$    &$1$  &$3^*$    &$1$ &$1$  &$1$  &$3^*$    &$1$  &$3$ &$1$ &$1$   &$1$     \\ \hline
$SU(2)_W$&$2$ &$1$      &$1$  &$1$      &$2$ &$1$  &$2$  &$1$      &$2$  &$1$ &$1$ &$1$   &$1$     \\ \hline
$y=6Y$   &$1$ &$-4$     &$6$  &$2$      &$-3$&$0$  &$-3$ &$2$      &$3$  &$-2$&$0$ &$0$   &$0$     \\ \hline  
$x$      &$1$ &$1$      &$1$  &$2$      &$2$ &$0$  &$-3$ &$-3$     &$-2$ &$-2$&$5$ &$0$   &$0$     \\ \hline
$z$      &$-1$&$-1$     &$-1$ &$2$      &$2$ &$-4$ &$-1$ &$-1$     &$2$  &$2$ &$-1$&$8$   &$-8$    \\ \hline
$R$      &$-$ &$-$      &$-$  &$-$      &$-$ &$-$  &$+$  &$+$      &$+$  &$+$ &$+$ &$+$   &$+$     \\ \hline
\end{tabular}
\end{center}
\caption{$G_2$ assignment of superfields. 
Where the $x$, $y$ and $z$ are charges of $U(1)_X$, $U(1)_Y$ and $U(1)_Z$, and $Y$ is hypercharge.}
\begin{center}
\begin{tabular}{|c|c|c|c|c|c|c|c|c|c|c|c|c|c|c|c|c|}
\hline
     &$Q_i$    &$Q_3$    &$U^c_i$  &$U^c_3$  &$E^c_1$  &$E^c_2$  &$E^c_3$   &$D^c_i$  &$D^c_3$  &$L_i$    &$L_3$    
     &$N^c_i$  &$N^c_3$  &$\Phi_i$ &$\Phi_3$ &$\Phi^c_a$ \\
      \hline 
$S_4$&${\bf 2}$&${\bf 1}$&${\bf 2}$&${\bf 1}$&${\bf 1}$&${\bf 1}$&${\bf 1'}$&${\bf 2}$&${\bf 1}$&${\bf 2}$&${\bf 1}$
     &${\bf 2}$&${\bf 1}$&${\bf 2}$&${\bf 1}$&${\bf 3}$ \\ 
	  \hline 
$Z_2$&$+$      &$-$      &$-$      &$+$      &$+$      &$-$      &$+$       &$-$      &$+$      &$+$      &$+$      
     &$+$      &$+$      &$-$      &$+$      &$+$ \\
      \hline
\end{tabular}
\end{center}
\caption{$S_4\times Z_2$ assignment of quark, lepton and flavon superfields in the model A (Where the index $i$ of 
the $S_4$ doublets runs $i=1,2$, and the index $a$ of the $S_4$ triplets runs $a=1,2,3$.).}
\begin{center}
\begin{tabular}{|c|c|c|c|c|c|c|c|c|c|c|c|c|c|c|c|}
\hline
     &$Q_i$    &$Q_3$    &$U^c_i$  &$U^c_3$  &$E^c_i$  &$E^c_3$  &$D^c_i$  &$D^c_3$                
     &$L_i$    &$L_3$    &$N^c_i$  &$N^c_3$  &$\Phi_i$ &$\Phi_3$ &$\Phi^c_a$ \\ \hline
$S_4$&${\bf 2}$&${\bf 1}$&${\bf 2}$&${\bf 1}$&${\bf 2}$&${\bf 1}$&${\bf 2}$&${\bf 1}$
     &${\bf 2}$&${\bf 1}$&${\bf 2}$&${\bf 1}$&${\bf 2}$&${\bf 1}$&${\bf 3}$  \\ \hline
$Z_2$&$+$      &$+$      &$+$      &$+$      &$-$      &$+$      &$+$      &$+$
     &$+$      &$-$      &$+$      &$+$      &$+$      &$-$      &$+$ \\ \hline
\end{tabular}
\end{center}
\caption{$S_4\times Z_2$ assignment of quark, lepton and flavon superfields in the model B (Where the notation is the same as that in Table 2.).}
\begin{center}
\begin{tabular}{|c|c|c|c|c|c|c|c|c|}
\hline      
     &$H^D_i$  &$H^D_3$  &$H^U_i$  &$H^U_3$  &$S_i$    &$S_3$    &$g_a$     &$g^c_a$  \\
      \hline
$S_4$&${\bf 2}$&${\bf 1}$&${\bf 2}$&${\bf 1}$&${\bf 2}$&${\bf 1}$&${\bf 3}$ &${\bf 3}$\\
      \hline
$Z_2$&$+$      &$-$      &$+$      &$-$      &$-$      &$+$      &$+$       &$+$      \\
      \hline
\end{tabular}
\end{center}
\caption{$S_4\times Z_2$ assignment of Higgs and g-quark superfields in the model A and model B (Where the notation is the same as that in Table 2.).}
\end{table}




\begin{thebibliography}{99}
\bibitem{SUSY}H.P. Nilles,  \PRP\vol{110}{1984}{1}.
\bibitem{extra-u1}F. Zwirner,  \IJMP\vol{A3}{1988}{49},
J.L. Hewett and T.G. Rizzo,  \PRP\vol{183}{1989}{193}.
\bibitem{mu-problem}D. Suematsu and Y. Yamagishi,  \IJMP\vol{A10}{1995}{4521}. 
\bibitem{f-extra-u1}R. Howl and S.F. King, JHEP\vol{0805}{2008}{008} [arXiv:0802.1909[hep-ph]].
\bibitem{review}
For reviews and recent advanced works, see the followings:
  P.~O.~Ludl,
  arXiv:0907.5587 [hep-ph].
  H.~Ishimori, T.~Kobayashi, H.~Ohki, H.~Okada, Y.~Shimizu and M.~Tanimoto,
  arXiv:1003.3552 [hep-th],
  L.~Merlo,
  arXiv:1004.2211 [hep-ph],
  W.~Grimus and P.~O.~Ludl,
  arXiv:1006.0098 [hep-ph],
  K.~S.~Babu and S.~Gabriel,
  arXiv:1006.0203 [hep-ph],
  P.~O.~Ludl,
  arXiv:1006.1479 [math-ph].

\bibitem{s4}S. Pakvasa and H. Sugawara,  \PL\vol{B73}{1978}{61},
E. Ma, \PL\vol{B632}{2006}{352} [arXiv:hep-ph/0508231],
C. Hagedorn, M. Lindner and R.N. Mohapatra, JHEP\vol{0606}{2006}{042}
[arXiv:hep-ph/0602244],
 H.~Zhang,
  Phys.\ Lett.\  B {\bf 655}, 132 (2007)
  [arXiv:hep-ph/0612214],
Y. Koide, JHEP\vol{0708}{2007}{086}
[arXiv:0705.2275 [hep-ph]],
F. Bazzocchi and S. Morisi, [arXiv:0811.0345 [hep-ph]],
H.~Ishimori, Y.~Shimizu and M.~Tanimoto,
  Prog.\ Theor.\ Phys.\  {\bf 121}, 769 (2009)
  [arXiv:0812.5031 [hep-ph]],
F.~Bazzocchi, L.~Merlo and S.~Morisi,
  Nucl.\ Phys.\  B {\bf 816} (2009) 204
  [arXiv:0901.2086 [hep-ph]],
F.~Bazzocchi, L.~Merlo and S.~Morisi,
  Phys.\ Rev.\  D {\bf 80} (2009) 053003
  [arXiv:0902.2849 [hep-ph]],
   G.~Altarelli, F.~Feruglio and L.~Merlo,
  JHEP {\bf 0905} (2009) 020
  [arXiv:0903.1940 [hep-ph]],
  W.~Grimus, L.~Lavoura and P.~O.~Ludl,
  J.\ Phys.\ G {\bf 36}, 115007 (2009)
  [arXiv:0906.2689 [hep-ph]],
  G.~J.~Ding,
  Nucl.\ Phys.\  B {\bf 827}, 82 (2010)
  [arXiv:0909.2210 [hep-ph]],
   B.~Dutta, Y.~Mimura and R.~N.~Mohapatra,
  arXiv:0911.2242 [hep-ph],
  D.~Meloni,
  J.\ Phys.\ G {\bf 37}, 055201 (2010)
  [arXiv:0911.3591 [hep-ph]],
  G.~J.~Ding and J.~F.~Liu,
  JHEP {\bf 1005}, 029 (2010)
  [arXiv:0911.4799 [hep-ph]],
  S.~Morisi and E.~Peinado,
  Phys.\ Rev.\  D {\bf 81}, 085015 (2010)
  [arXiv:1001.2265 [hep-ph]],
  C.~Hagedorn, S.~F.~King and C.~Luhn,
  arXiv:1003.4249 [hep-ph],
  Y.~H.~Ahn, S.~K.~Kang, C.~S.~Kim and T.~P.~Nguyen,
  arXiv:1004.3469 [hep-ph],
  H.~Ishimori, K.~Saga, Y.~Shimizu and M.~Tanimoto,
  arXiv:1004.5004 [hep-ph].


\bibitem{a4}E.Ma and G.Rajasekaran, \PR\vol{D64}{2001}{113012} [arXiv:hep-ph/0106291],
K.S. Babu, E. Ma and J.W.F. Valle, \PL\vol{B552}{2003}{207} [arXiv:hep-ph/0206292],
G. Altarelli and F. Feruglio, \NP\vol{B741}{2006}{215} [arXiv:hep-ph/0512103],
X.G. He, Y.Y. Keum and R.R. Volkas, JHEP\vol{0604}{2006}{039} [arXiv:hep-ph/0601001],
M. Honda and M. Tanimoto, \PTP\vol{119}{2008}{583} [arXiv:0801.0181 [hep-ph]].
\bibitem{t-prime}A. Aranda, C.D. Carone and R.F. Lebed, \PR\vol{D62}{2000}{016009} [arXiv:hep-ph/0002044],
P.H. Frampton and T.W. Kephart, JHEP\vol{0709}{2007}{110} 
[arXiv:0706.1186 [hep-ph]],
S. Sen, \PR\vol{D76}{2007}{115020} [arXiv:0710.2734 [hep-ph]],
G.J. Ding, \PR\vol{D78}{2008}{036011} 
[arXiv:0803.2278 [hep-ph]],
M.C. Chen, K.T. Mahanthappa and F. Yu, arXiv:0907.3963 [hep-ph],
 F.~Feruglio, C.~Hagedorn, Y.~Lin and L.~Merlo,
  Nucl.\ Phys.\  B {\bf 775} (2007) 120
  [arXiv:hep-ph/0702194].
\bibitem{d27}I.de Medeiros Varzielas, S.F. King and G.G. Ross,
\PL\vol{B648}{2007}{201} [arXiv:hep-ph/0607045],
E. Ma, \MPL\vol{A21}{2006}{1917} [arXiv:hep-ph/0607056].
\bibitem{d54}H. Ishimori, T. Kobayashi, H. Okada, Y. Shimizu and M. Tanimoto, JHEP \vol{0904}{2009}{011} [arXiv:0811.4683 [hep-ph]],
H. Ishimori, T. Kobayashi, H. Okada, Y. Shimizu and M. Tanimoto, arXiv:0907.2006 [hep-ph].
\bibitem{lepton}J. Kubo, \PL\vol{B578}{2004}{156}.
\bibitem{higgs}
  J.~Kubo, H.~Okada and F.~Sakamaki, Phys.\ Rev.\  D {\bf 70}, 036007 (2004)
  [arXiv:hep-ph/0402089].
\bibitem{mass}Z. Xing, H. Zhang and Z. Zhou, [arXiv:0712.1419[hep-ph]].
\bibitem{PDG2008}Particle Date Group \PL\vol{B667}{2008}{1}.
\end{thebibliography}
\end{document}